\documentclass{article}
\usepackage{spconf,amsmath,graphicx,hyperref,xcolor,tabularray,comment}

\definecolor{bestgreen}{RGB}{0,128,96}
\definecolor{belowred}{RGB}{190,55,45}
\newcommand{\best}[1]{\textcolor{bestgreen}{#1}}
\newcommand{\below}[1]{\textcolor{belowred}{#1}}
\newif\ifshowfigures
\showfigurestrue
\newif\ifshowexamfigure
\showexamfiguretrue

\title{MIXJUDGE: LEARNING A LANGUAGE-QUERYABLE CRITIC FOR VOCAL MIXING}

\name{Kuan-Hsun Ho$^{1}$, Yuan-Chiao Cheng$^{2}$, Yen-Tung Yeh$^{3}$, Yi-Hsuan Yang$^{3,4}$}
\address{$^{1}$Independent Researcher\\
$^{2}$Music Informatics Group, Georgia Institute of Technology\\
$^{3}$Music and AI Lab, National Taiwan University\\
$^{4}$AI-CoRE, National Taiwan University
}

\begin{document}
\ninept
\maketitle

\begin{abstract}
Mixing requires not only applying processing but also assessing the resulting 
mix. Audio-language models can provide conversational mixing advice, but 
existing evaluations do not separately test whether they can judge
specific mixing issues. Since professional supervision for such judgments
is scarce, we construct a curriculum of 11 vocal-mix issues at
multiple severity levels and use it to train MixJudge, a language-queryable
audio critic. MixJudge shows its ability to retrieve mixing issues, distinguish
opposite issues, and order the severity of an issue.
Moreover, its judgments transfer to menu selection: given a set of 
processing candidates, MixJudge often favors versions acoustically closer 
to a held-out professional version, demonstrating how explicit mixing 
judgment can support comparison and downstream mixing decisions.
\end{abstract}


\section{Introduction}
\label{sec:intro}

Mixing is a central stage of music production in which engineers process and
combine recorded tracks into a coherent whole \cite{deman2019intelligent}.
Balancing level, tone, dynamics, space, and source interaction requires domain
expertise, and the appropriate treatment of a source depends on the
surrounding arrangement \cite{martinez2022ood,yeh2026stemblending}. Mixing is
also not a single-answer problem: engineers working with the same material
still produce substantially different results
\cite{deman2014features,deman2017dataset}. In practice, demo mixes and reference
songs are used to communicate and establish expectations for the desired
outcome \cite{vanka2024communication}. These properties make mixing
not only a problem of producing a treatment, but also one of assessing the
resulting mix state.

Existing automatic mixing research primarily addresses the production side of
this problem. Reference-free systems infer processing directly from source
stems (isolated tracks) \cite{martinez2022ood,yeh2026stemblending,steinmetz2021mixing,
moliner2025megami}, while reference-conditioned mixing style transfer uses an
example recording to communicate desired production characteristics
\cite{koo2023style,vanka2024diffmst}. Both settings focus on producing a mixing 
outcome from the source material and, optionally, an audio reference, rather than 
separately assessing the resulting mix.

Audio-language research provides natural-language interfaces for reasoning
about audio. Existing work supports audio-text representation learning
\cite{elizalde2023clap}, text-guided editing and effect control
\cite{chu2025text2fx,copet2024instructmusicgen}, and executable effect chains
\cite{koo2026llm2fx}. Closest to our setting, MixAssist fine-tunes
audio-language models to generate audio-grounded conversational mixing advice
and evaluates the usefulness of the generated advice as a whole
\cite{clemens2025mixassist}. This establishes conversational
mixing assistance, but does not separately examine whether a model can
make issue-specific audio judgments that vary consistently with the severity of
an issue.

We study a complementary role: an explicit critic that assesses a mix state
with respect to a language-specified issue. 
Applying the same judgment to different versions of a mix then provides
a common basis for comparing processing alternatives.
Such judgment depends on an experience-shaped
expectation of what constitutes a plausible mix outcome for the song, which we
call a mixing prior. Language then enables an engineer to ask whether a mix is
\emph{too loud}, \emph{dull}, or affected by another issue rather than receiving
only an opaque quality score.

Training data for this kind of judgment are scarce. Public datasets rarely
contain both unprocessed multitracks and their professional mixes, and multiple
professional realizations of the same material are rarer still
\cite{martinez2022ood,moliner2025megami,koo2023style}. We instead synthesize
known mixing issues at multiple severity levels and pair them with text,
allowing a model to contrast a clean clip with its degraded versions while
musical content remains fixed. To keep the synthetic degradations controlled
while preserving a meaningful source-context relationship, we focus on vocals
and their musical backing (sum of all non-vocal sources). We
present MixJudge, a language-queryable audio critic that learns from this
curriculum to assess whether a vocal mix exhibits a language-specified issue.

We evaluate MixJudge in two stages: a critic’s exam of its mixing judgments, followed 
by menu selection to test their use in a downstream mixing decision.
This work makes three contributions: 1) a vocal-mix curriculum providing paired
audio-text supervision over 11 issues and multiple severity levels; 2) MixJudge, 
a language-queryable critic that judges whether a vocal mix exhibits a specified 
issue; and 3) menu selection, a concrete comparison-based form of mixing
assistance showing how these judgments can support decisions.

\section{Vocal-Mix Curriculum}
\label{sec:curriculum}

\begin{comment}
We construct the curriculum through audio-caption synthesis that creates
known mixing states.
\end{comment}
Here, a \emph{clip} denotes a 10-second
vocal-and-backing excerpt, an \emph{issue} names an audible
problem, a \emph{recipe} is signal processing that instantiates an issue, and a
\emph{state} is one processed version of a clip.

The curriculum covers 11 issue labels plus \emph{clean}, where the issues form a
teaching vocabulary rather than an exhaustive account of mixing quality, and
\emph{clean} denotes no injected problem rather than a universally optimal mix. The
issue labels are \emph{too loud}, \emph{too quiet}, \emph{muddy},
\emph{thin}, \emph{harsh}, \emph{dull}, \emph{too wet} (excessive reverberation), 
\emph{too dry} (insufficient reverberation),
\emph{over-compressed}, \emph{under-compressed}, and \emph{masking}
(the backing obscures the vocal). The first
10 labels form 5 opposite pairs. \emph{Masking} has no explicitly named opposite
in text vocabulary, so we do not force a twelfth symmetric label. 
\begin{comment}
Interestingly, our
preliminary analysis of practitioner language found that the reverse
interaction, in which the vocal masks the backing, is usually described as
\emph{muddy}. 
\end{comment}
In general, the
11 issue labels use common mixing terminology and cover level, tone, space, dynamics, and source interaction,
reflecting the perceptual features studied in multitrack music mixtures
\cite{deman2014features}.

We keep the audio states mono to limit construction and modeling complexity,
thus excluding stereo-width issues. Phase is also omitted because the available public
stems do not preserve the multi-microphone relationships required to construct realistic 
phase problems.

\begin{comment}
\subsection{Audio and caption synthesis}
\end{comment}

\noindent\textbf{Audio and caption synthesis.}
Each recipe follows three stages. First, an eligibility check determines
whether the clip contains the audio material needed to express the issue.
Second, the recipe applies a degradation at one of three severity levels: mild, medium, or severe.
Then, most non-level recipes normalize the loudness of the affected source 
so that the intended issue is not confounded by an overall level change.
Third, recipe-specific acceptance rules retain only states
that satisfy recipe-specific acoustic checks, proxy thresholds, and safety conditions. These
rules provide implementation-level filtering rather than perceptual
validation. The curriculum contains 18 heuristic recipes; some issues
use multiple recipes so that one issue is not tied to only one signal
transformation.

Each retained state receives a template-generated caption following
``The [subject] [is/sounds] [optional degree] [quality] [optional scope].''
After the issue and severity are chosen, the generator samples
subjects, degree words, issue phrases, and scopes from a fixed 
lexicon.\footnote{Code, recipes, and caption-generation details are available at
\url{https://github.com/jasonho610/mixjudge-esntl}.} Taking
recipe \emph{muddy tone} as an example, the recipe 
requires the vocal's low-mid energy ratio to exceed an eligibility threshold,
then boosts the low-mid region according to severity, and normalizes the
loudness of the affected source. 
One accepted severe state may pair this audio
with ``The [voice] [sounds] [completely] [muddy] [in the low mids].''

\section{MixJudge}
\label{sec:method}

MixJudge comprises CLAP audio and text encoders \cite{elizalde2023clap}, 
learned role embeddings,
and a relational fusion network. For a state $A=(b,v,m)$, it receives
backing $b$, vocal $v$, their blend $m$, and a text description $q$ naming the issue
to judge. The audio encoder maps the three audio inputs to $h_b$, $h_v$, and
$h_m$. A fusion network $F$ then forms a state embedding
\begin{equation}
 z_A=\operatorname{norm}F([\tilde h_b,\tilde h_v,\tilde h_m,
 \tilde h_v-\tilde h_b,\tilde h_m-\tilde h_b]),
 \label{eq:fusion}
\end{equation}
where $\operatorname{norm}(x)=x/\lVert x\rVert_2$ and $\tilde h_r=h_r+e_r$ 
with $e_r$ denoting a learnable role embedding.
The intuition for forming role-marked slots and difference features is to
expose how the vocal and resulting blend relate to the backing. With text
encoder mapping $q$ to a normalized embedding $z_q$,
$s(A,q)=z_A^\top z_q$ formulates the cosine similarity between the audio state
and the issue description.

\subsection{Training}

\ifshowfigures
\begin{figure}[t]
\centering
\includegraphics[width=0.9\columnwidth]{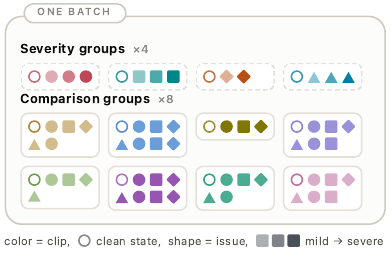}
\caption{Same-content training batch construction.}
\label{fig:batches}
\end{figure}
\else
\refstepcounter{figure}\label{fig:batches}
\fi

To prevent the model from using song or singer identity as a shortcut, every
training group contains different states of the same clip. A severity group
contains the clean state and every available severity of one recipe, providing
severity-order supervision. On the other hand, a comparison group contains
the clean state and up to seven recipes at the same nominal severity,
providing issue discrimination (Fig.~\ref{fig:batches}).

Training combines state matching, two auxiliary margin losses, and severity
ranking. State-level matching uses a pairwise sigmoid loss $\mathcal L_{\rm sig}$ following SigLIP
\cite{zhai2023siglip}. Unlike diagonal-only InfoNCE \cite{oord2018cpc}, it can
mark multiple audio-caption pairs as positive.
Audio state and caption pairs sharing the same recipe and nominal severity 
are positive, including pairs from different clips; all other pairs are negative. 
This dense supervision encourages audio-text
correspondence for curriculum states while reducing reliance on
musical content.

To enhance semantic discrimination among issues, we introduce two
auxiliary margin losses: the issue-selectivity loss
$\mathcal L_{\rm sel}$ and the opposite-issue loss
$\mathcal L_{\rm opp}$. Since batch-sampled captions vary and do not 
guarantee coverage of every issue,
both losses use one fixed issue anchor $q_i$ 
per issue as a stable comparison axis. Each anchor also follows the caption 
template without a degree modifier. Let $A_i^s$ denote a clip rendered with
issue $i$ at severity $s$, and let $A_i^0\equiv A^0$ denote its clean state.
For the issue-selectivity loss, define the clean-relative score change
$\Delta_j=s(A_i^s,q_j)-s(A^0,q_j)$, then
\begin{equation}
\mathcal L_{\rm sel}=\operatorname{ReLU}
 [0.05-\Delta_i+\max_{j\ne i}\Delta_j].
\end{equation}
For the opposite-issue loss, let $q_{\bar i}$ name the opposite of issue $i$:
\begin{equation}
\mathcal L_{\rm opp}=\operatorname{ReLU}
 [0.10-s(A_i^s,q_i)+s(A_i^s,q_{\bar i})].
\end{equation}
\emph{Masking} is omitted from this loss due to its lack of the opposite label.

As for severity-order supervision, the ranking loss $\mathcal L_{\rm rank}$ 
\cite{burges2005ranknet} uses every ordered pair in a severity group,
including clean-to-degraded comparisons:
\begin{equation}
\mathcal L_{\rm rank}=\operatorname{softplus}
 [-\{s(A_i^{s_2},q_i)-s(A_i^{s_1},q_i)\}],\qquad s_2>s_1.
\end{equation}
Finally, the full objective is
\begin{equation}
 \mathcal L=\mathcal L_{\rm sig}+.3\mathcal L_{\rm rank}
 +.1\mathcal L_{\rm sel}+.1\mathcal L_{\rm opp}, 
 \label{eq:objective}
\end{equation}
where the margins and weights were set heuristically.

\noindent\textbf{Implementation Details.}
The fusion network maps its 2,560-dimensional input through 
2,048- and 1,024-unit GELU layers with 0.1 dropout to 512 dimensions. We train for 60 
epochs with AdamW (weight decay \(10^{-2}\)), using learning rates 
\(3\times10^{-4}\) for fusion and \(3\times10^{-5}\) for adapted encoders. 
We report MixJudge with both CLAP encoders frozen (fusion-only) and with its audio and text encoders also adapted (adapted). Fusion-only 
updates the fusion network, role embeddings, and SigLIP scale/bias; adapted additionally 
updates the final HTSAT stage, final two RoBERTa layers, and their projections.

\section{Experimental Setup}
\label{sec:experiments}

We evaluate whether MixJudge acquires the issue-level judgments targeted by the 
curriculum through the critic’s exam, and whether those judgments support a 
downstream mixing decision through menu selection.

For training, data synthesis yields 78,740 state-caption examples from 236
MoisesDB \cite{pereira2023moisesdb} and 134 MUSDB18-HQ
\cite{stoter2018sisec} songs.
For evaluation, we construct 4,557 synthetic examples and 88
raw/professionally processed vocal pairs from 23 selected songs drawn from the
combined MedleyDB 1.0 and 2.0 releases
\cite{bittner2014medleydb,bittner2016medleydb2}.
We exclude classical songs, songs
with multiple vocal stems or fewer than three usable stems, and songs failing
the vocal-audibility floor.
The training set excludes tracks shared with the MedleyDB evaluation set.
All audio is sampled at 44.1 kHz.

\subsection{Baselines}

We compare against frozen CLAP as a joint audio-text embedding baseline, and against Music 
Flamingo \cite{ghosh2025musicflamingo}, Audio Flamingo Next \cite{ghosh2026afnext}, 
and Qwen2.5-Omni \cite{xu2025qwen25omni} as generative audio-language model baselines. Following AQAScore
\cite{kuan2026aqascore}, we ask each generative model a fixed yes/no question
for every issue and use the YES–NO logit margin as its score. 
The compared systems receive only the blend, whereas MixJudge receives
the vocal, backing, and blend. Every applicable system is evaluated on identical
candidates and pools, although its architecture and scoring
interface remain native to that system.
We also include Chance, the expected performance of random ranking or
selection given the same evaluation inputs.

% Declare result floats before the evaluation details so both tables can
% appear on page 4 while the Results discussion remains in source order.
\begin{table*}[t]
\centering
\caption{Results of the critic's exam and menu selection. Retrieval entries
report R@1/MRR. $A_{\rm opp}$ is both-poles accuracy; $\rho_{\rm sev}$,
severity correlation; $A_{\rm chain}$, full-order accuracy; $S_{\rm sel}$,
selectivity; $C$, closer rate; $\rho_{\rm menu}$, whole-menu correlation; and
$R_{\rm med}$, median recovery. Slash-separated menu entries correspond to
mean/max evidence policies. Green marks the best result; red marks performance
below chance.}
\label{tab:main-results}
\begin{tblr}{
  colsep=0pt,
  colspec={Q[l,m,wd=40mm]*{2}{Q[c,m,wd=15.5mm]}
           Q[c,m,wd=20.5mm]*{3}{Q[c,m,wd=9.2mm]}*{3}{Q[c,m,wd=17.7mm]}},
  column{1}={l,m},
  column{2-Z}={c,m},
  cell{1}{1}={r=3}{m},
  cell{1}{2}={c=6}{c},
  cell{1}{8}={c=3,r=2}{c},
  cell{2}{2}={c=2}{c},
  cell{2}{5}={c=3}{c},
  row{10}={font=\itshape},
  vline{2}={1-10}{solid},
  vline{4,5}={2-10}{dashed},
  vline{8}={1-10}{solid},
  hline{1,11}={-}{0.08em},
  hline{2}={2-7}{dashed},
  hline{3}={2-10}{},
  hline{4,10}={-}{},
  hline{5,8}={-}{dashed},
}
System              & {The critic's exam} & & & & & & {Menu selection} & & \\
                    & {Issue retrieval} & & {Opposite issue} & {Severity ordering} & & & & & \\
                    & $T\!\rightarrow\!A$ & $A\!\rightarrow\!T$ & $A_{\rm opp}$ & $\rho_{\rm sev}$ & $A_{\rm chain}$ & $S_{\rm sel}$ & $C$ & $\rho_{\rm menu}$ & $R_{\rm med}$ \\
Frozen CLAP         & \below{.050/.203} & .066/.230 & .266 & .331 & .442 & .172 & \below{.330/.284} & .146/.147 & \below{$-$.552/$-$.636} \\
Music Flamingo      & .097/.255 & \below{.061/.212} & .250 & .235 & .189 & .119 & \below{.364/.318} & .168/.056 & $-$.218/\below{$-$.303} \\
Audio Flamingo Next & .093/.237 & \below{.041/.191} & \below{.243} & .175 & .170 & .129 & .523/.386 & .350/.082 & \ \ .134/\below{$-$.466} \\
Qwen2.5-Omni        & .142/.296 & \below{.057/.203} & .374 & .309 & .239 & .197 & .398/.386 & .290/.305 & $-$.060/$-$.211 \\
MixJudge (ours, fusion-only) & .112/.274 & .097/.254 & .727 & .768 & .713 & .793 & .705/.648 & .609/.638 & .567/.500 \\
MixJudge (ours, adapted)  & \best{.164/.321} & \best{.117/.297} & \best{.839} & \best{.971} & \best{.922} & \best{.838} & \best{.727/.716} & \best{.730/.721} & \best{.665/.684} \\
Chance              & .064/.214 & .064/.214 & .250 & .000 & .048 & .091 & .380/.380 & .000/.000 & $-$.291/$-$.291
\end{tblr}
\end{table*}

\begin{table}[t]
\centering
\caption{Objective ablation.}
\label{tab:objective-ablation}
\begin{tblr}{
  colsep=.3pt,
  leftsep=0pt,
  rightsep=0pt,
  colspec={Q[l,m,wd=18mm]Q[c,m,wd=14mm]*{3}{Q[c,m,wd=8mm]}*{2}{Q[c,m,wd=14mm]}},
  column{1}={l,m},
  cell{1}{2}={c=4}{c},
  cell{1}{6}={c=2}{c},
  cell{1}{1}={r=2}{m},
  vline{2}={1-6}{solid},
  vline{6}={1-6}{solid},
  hline{1,7}={-}{0.08em},
  hline{2-3}={-}{},
  hline{4}={-}{dashed},
}
Objective & The critic's exam & & & & Menu selection & \\
& $T\!\rightarrow\!A$ & $A_{\rm opp}$ & $\rho_{\rm sev}$ & $S_{\rm sel}$ & $C$ & $\rho_{\rm menu}$ \\
InfoNCE
 & .144/.317 & .812 & .611 & .495 & .636/.682 & .501/.650 \\
$\mathcal L_{\rm sig}$
 & .119/.317 & .725 & .578 & .356 & \best{.739/.773} & .728/\best{.728} \\
\hspace{.5em}$+$ $\mathcal L_{\rm rank}$
 & .098/.259 & .786 & .916 & .384 & .705/.614 & .680/.558 \\
\hspace{1em}$+$ $\mathcal L_{\rm sel, opp}$
 & \best{.164/.321} & \best{.839} & \best{.971} & \best{.838} & .727/.716 & \best{.730}/.721 \\
\end{tblr}
\end{table}

\ifshowexamfigure
\begin{figure}[t]
\centering
\includegraphics[width=\columnwidth]{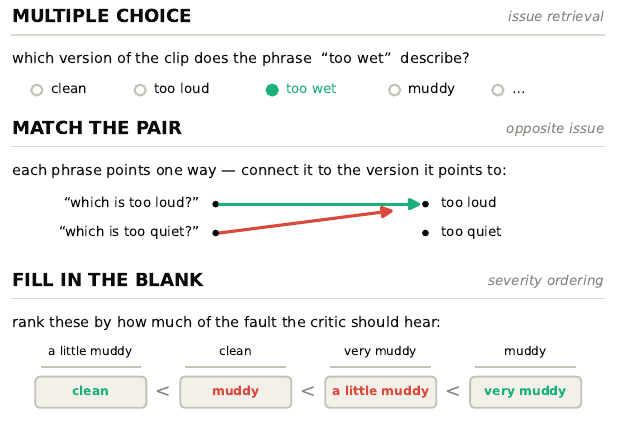}
\caption{The critic's exam. Green marks correct responses; red marks errors.}
\label{fig:critic-exam}
\end{figure}
\else
\fi

\subsection{The critic's exam}

The critic's exam comprises issue retrieval, opposite-issue discrimination, and severity
ordering (Fig.~\ref{fig:critic-exam}).

\noindent\textbf{Issue retrieval.} Given multiple states of one clip, a text description must
retrieve a state expressing its issue; any state with the queried issue is considered relevant, 
regardless of recipe or severity. Each description ranks a same-clip pool shared by all systems, while
chance is computed from the pool size and number of relevant states. The reverse
task retrieves the corresponding issue description from
audio. We report Recall@1 (R@1) and mean reciprocal rank (MRR) in both 
text-to-audio ($T\!\rightarrow\!A$) and audio-to-text ($A\!\rightarrow\!T$) 
retrieval directions.

\noindent\textbf{Opposite-issue discrimination.} For an opposite issue pair, each text description makes a
two-way choice between the corresponding
states. Both-poles accuracy $A_{\rm opp}$ counts a pair as correct only when both queries
choose correctly. 

\noindent\textbf{Severity ordering.} Within each clip, the queried issue score should
rise from clean state through increasing degradation strengths.
We report three metrics: 1) severity correlation $\rho_{\rm sev}$, Spearman's
$\rho$ with the known order; 2) full-order accuracy $A_{\rm chain}$, the
proportion of complete clean-to-severe sequences ordered correctly; and 3)
selectivity $S_{\rm sel}$, the proportion of severity sequences whose injected
issue shows the largest standardized clean-to-severe increase among all 11
issue scores.

\ifshowfigures
\begin{figure}[t]
\centering
\includegraphics[width=\columnwidth]{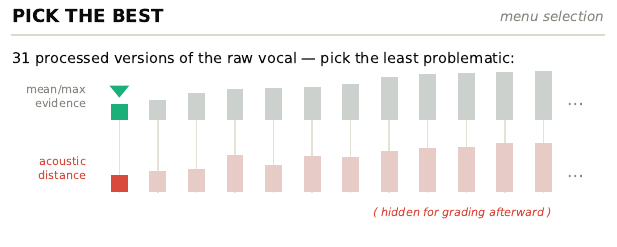}
\caption{Menu selection. Gray shows issue evidence; red shows each candidate’s $D$ to professional version.}
\label{fig:menu-selection}
\end{figure}
\else
\refstepcounter{figure}\label{fig:menu-selection}
\fi

\subsection{Menu selection}

Menu selection tests one downstream use of MixJudge's issue scores in a
concrete mixing decision (Fig.~\ref{fig:menu-selection}). For each raw vocal, 
we render the same fixed menu of 31 processing candidates: no edit, 18 single-control
settings spanning gain, low- and mid-band equalization, compression, and
reverb, and 12 two-control combinations drawn from the same operations. Thus,
each candidate is the audible outcome of a particular processing treatment 
and is evaluated against the same backing.

For candidate $A$, let $q_0$ denote the clean-state description. The evidence
for issue $i$ is $e_i(A)=s(A,q_i)-s(A,q_0)$. We form two selection policies:
mean evidence summarizes the 11 evidence values, while maximum evidence uses
the strongest issue evidence. Each policy selects the candidate with the lowest
aggregate value.
Only after selection do we reveal the professionally processed vocal. We measure
its acoustic distance $D$ from each candidate using the absolute RMS-level
difference plus the mean absolute log-energy difference across eight octave
bands from 62~Hz to 16~kHz.
Although the professional version is one realization rather than a uniquely
best mix, approaching it tests whether the learned issue evidence transfers
beyond synthesized curriculum states. We report three metrics: 1) closer rate
$C$, the proportion of selections having $D$ lower than that of no edit; 2) whole-menu
correlation $\rho_{\rm menu}$, Spearman's $\rho$ between ascending aggregated
issue evidence and ascending $D$; and 3) median recovery $R_{\rm med}$,
computed over per-clip values,
\begin{equation}
R=(D_0-D_{\rm pick})/(D_0-D_{\rm oracle}),
\end{equation}
where $D_0$ is no-edit distance
and $D_{\rm oracle}$ is the minimum distance among candidates. Here, $R=0$ denotes
no improvement, $R=1$ reaches the menu oracle, and $R<0$ moves away from the
professional version.

\section{Results}
\label{sec:results}

\subsection{What MixJudge learns}

Table~\ref{tab:main-results} shows that the baselines exhibit some
issue-retrieval ability, but their opposite-issue discrimination and
severity-ordering measures remain well below MixJudge. For the generative
audio-language models, text-to-audio retrieval is also consistently stronger
than audio-to-text retrieval. This asymmetry is
consistent with recent findings of text bias in large audio-language models,
although our evaluation does not isolate its cause \cite{wang2025textbias}.
Adapted MixJudge achieves the highest values across every reported critic-exam
metric, showing that it acquired the abilities targeted by the curriculum and
training objectives. This does not establish that the tested large
audio-language models lack mixing knowledge; rather, their scores under this
interface do not reliably solve the individual critic tasks. 
This also supports the need for evaluating mixing judgment through separate 
capabilities using defined evaluators instead of a single holistic score 
\cite{li2026anyaudiojudge}.

\subsection{From known issues to mixing assistance}

As shown in the menu-selection columns of Table~\ref{tab:main-results}, MixJudge performs 
strongly on three criteria: selecting a candidate with distance smaller than $D_0$, ranking candidates 
consistently with their $D$,
and recovering a substantial fraction of the improvement 
available on the menu. Together, these results suggest 
that the learned judgment is not limited to recognizing synthetic degradations: 
its evidence remains informative for ordering processing outcomes relative to 
a held-out professional version.
\begin{comment}
Rather than producing a treatment itself, 
the critic can compare candidate outcomes to support a mixing decision.
\end{comment}

We further audit whether this result is explained by the candidate menu or a
fixed processing shortcut. An improving candidate exists in 83 of 88 menus.
Conditioned on these 83 actionable menus, MixJudge selects an improving candidate
in 77.1\% and 75.9\% of cases using mean and max evidence policies, respectively. Its 
choices span 22 of the 31 candidates, showing that the
result is distributed across processing alternatives instead of being concentrated
on one treatment. 
\begin{comment}
However, MixJudge never selects no edit,
including for the five menus without an improving candidate. It demonstrates
useful candidate selection when an opportunity exists, but calibrated
abstention remains unresolved and is left for future work.
\end{comment}

\subsection{What each objective contributes}

Table~\ref{tab:objective-ablation} shows that the training objectives
contribute complementary properties rather than improving all metrics
uniformly. Overall, 
critic's exam and downstream selection are related, but improvements in one do
not translate monotonically to the other.
InfoNCE performs better than plain \(\mathcal L_{\rm sig}\) on most
critic-exam measures, whereas \(\mathcal L_{\rm sig}\) transfers better to menu
selection. Yet weaker opposite-issue discrimination, severity ordering,
and issue selectivity make strong downstream performance difficult to
attribute to the intended issue semantics. Adding
\(\mathcal L_{\rm rank}\) substantially improves severity ordering, but leaves
poor issue selectivity and inferior menu selection. The $\mathcal L_{\rm sel}$ and
$\mathcal L_{\rm opp}$ then sharply improve all critic-exam metrics,
while recovering strong menu correlation. The full objective makes the scores
more consistent with the intended issue identity and severity structure while
retaining strong downstream utility.

\section{Conclusion}
\label{sec:conclusion}

MixJudge shows that a vocal-mix curriculum can train a
language-queryable critic to make explicit mixing judgments. On the critic’s 
exam, MixJudge recognizes queried issues, discriminates opposite issues, 
and orders issue severity, while menu selection shows that its issue 
evidence can transfer beyond synthetic degradations to support decisions.
Together, these results establish explicit judgment and comparison
as a complementary form of language-assisted mixing. Future work can expand
the curriculum to broader mixing issues, develop MixJudge as an interactive
language-queryable critic, or extend menu selection into a practical mixing
workflow, where treatments can be proposed, their outcomes judged, and the 
proposals iteratively refined.

\bibliographystyle{IEEEbib}
\bibliography{mixjudge_refs}

\end{document}